%% file: main.tex
\documentclass[pdflatex,sn-basic,Numbered]{sn-jnl}

\usepackage[T1]{fontenc}
\usepackage{lmodern}
\usepackage{microtype}
\usepackage{amsmath,amssymb}
\usepackage{graphicx}
\usepackage{subcaption}
\usepackage{booktabs}
\usepackage{siunitx}
\usepackage{array}
\usepackage[capitalize,nameinlink,noabbrev]{cleveref}

\graphicspath{{figures/}}

\newcommand{\Rc}{R_c}

\newcommand{\Pran}{\mathrm{Pr}}
\newcommand{\Rrho}{R_\rho}

\newcommand{\RouteSurvivalTitle}{Route survival and spectral modification of finite-depth salt-finger plume forests under imposed mean shear}
\newcommand{\RouteSurvivalShortTitle}{Route survival of finite-depth salt-finger plume forests}
\newcommand{\RouteSurvivalKeywords}{salt fingers; double diffusion; finite-depth mixing; mean shear; spectral memory; plume coherence}
\newcommand{\RouteSurvivalAbstractText}{Salt-finger plume forests in a finite layer can differ in strength and in the route by which interfacial activity becomes vertically connected. We use direct three-dimensional simulations to test whether such a route is a short-lived realization-specific transient or a persistent route family under an added mean-shear perturbation. The baseline route atlas holds density ratio, diffusivity ratio, Prandtl number, interface thickness, roughness amplitude, domain, and resolution fixed while varying the imposed interfacial roughness spectrum. Low-mode roughness forms a broad connecting endpoint, high-annulus roughness forms a localized route-memory endpoint, and mixed roughness forms a delayed scale-transfer route. A second mixed realization preserves continuous active-width, spectral, and transport measures after \(t=45\), with mean absolute differences of \(3.1\%\) in \(w\)-active width, \(1.6\%\) in salinity-active width, \(2.8\%\) in broad spectral fraction, and \(3.6\%\) in salt flux, while shifting the binary scalar-contact label. We then impose an initial tanh mean shear on the mixed route. The full-resolution shear case reaches \(t=60\) and preserves finite-depth reach: first velocity contact occurs at \(t=57.75\), first salinity contact occurs at \(t=59.5\), and both times match the unsheared mixed reference. The spectral branch is redistributed. At \(t=60\), the broad fraction is \(1.116\) times the mixed value, the intermediate fraction is \(0.530\) times the mixed value, and the short-wave fraction is \(1.278\) times the mixed value. In this finite-depth configuration, route survival means preserved reach and contact timing with a changed spectral pathway.}

\begin{document}
\articletype{Research Article}
\title[\RouteSurvivalShortTitle]{\RouteSurvivalTitle}
\author*[1]{\fnm{Sriram P.} \sur{Kalathoor}}\email{sriram2@gatech.edu}
\affil*[1]{\orgdiv{Daniel Guggenheim School of Aerospace Engineering}, \orgname{Georgia Institute of Technology}, \orgaddress{\city{Atlanta}, \state{Georgia}, \country{USA}}}
\abstract{\RouteSurvivalAbstractText}
\maketitle

\section*{Article Highlights}
\begin{itemize}
  \item Mean shear preserves late finite-depth reach while changing the spectral pathway.
  \item Low-mode, high-annulus, and mixed roughness define distinct plume-forest routes.
  \item Mixed-seed comparison separates route survival from threshold-sensitive contact labels.
\end{itemize}

\section{Introduction}

Salt fingering occurs when warm, salty fluid overlies cooler, fresher fluid, so that the destabilizing salinity contrast diffuses more slowly than the stabilizing temperature contrast. The mechanism is classical \citep{stern1960salt,turner1974double}, and remains central to double-diffusive convection theory \citep{radko2013double}. The nonlinear state that develops in a finite layer can extend beyond the local instability criterion. Finite-amplitude fingers have growth and length constraints \citep{schmitt1979growth,kunze1987limits}, can undergo collective and secondary instabilities \citep{stern1969collective,simeonov2009dns}, and can organize into larger-scale motions and staircases \citep{traxler2011dynamics,stellmach2011dynamics}. Oceanic observations and reviews further motivate the transport problem: salt fingering can contribute to measurable diapycnal exchange and tracer structure in thermohaline water columns \citep{schmitt1994double,schmitt2005enhanced}.

Many descriptions of fingering transport begin from local parameters such as \(\Rrho\), \(\Pran\), and \(\tau\). Those parameters are essential, while the finite-amplitude state of the interface remains an additional ingredient. In a finite layer, the interface can already carry long waves, annular roughness, mixed spectral content, or remnants of previous motion. The subsequent plume forest then inherits both the local double-diffusive instability and the geometry of the initial disturbance. The central question becomes how that inherited structure affects vertical reach and exchange after the fingers have entered a nonlinear finite-depth regime.

The usual local descriptors answer whether a background stratification is fingering-favorable. A finite interface also has geometry and memory. It can be corrugated by waves, strained by shear, disturbed by intrusions, or left rough by earlier mixing. Those finite-amplitude features leave the basic thermohaline mechanism intact while changing the pathway by which a plume forest develops. The pathway matters in a finite-depth system because two plume forests with comparable local instability may differ in whether they remain near the interface, transfer energy into broader modes, or reach remote layers.

We call that pathway a route. A route is a coupled state made from contact timing, active vertical width, spectral branch, scalar exchange, and persistence under perturbation. This definition separates physically distinct endpoint states, quantifies realization-to-realization variability, and identifies which part of the plume forest changes under perturbation. In the present study, those parts are finite-depth reach, contact timing, spectral partition, and exchange.

This paper asks whether the delayed mixed route in a finite-depth salt-finger configuration survives an imposed mean shear. The question is intentionally narrower than a universal shear law for salt fingers. The configuration, density ratio, diffusivity ratio, interface thickness, domain, and roughness amplitude are fixed. The no-shear simulations first define the route atlas: low-mode roughness gives a broad connecting endpoint, high-annulus roughness gives a localized route-memory endpoint, and mixed roughness gives a delayed route between them. A second mixed realization then defines the natural tolerance of that route family. Finally, a full-resolution mean-shear perturbation tests whether finite-depth reach, transport, and spectral branch respond together or separately.

The no-shear route atlas extends the finite-depth roughness-spectrum comparison in \citet{kalathoor2026roughnessPreprint}, while the route-memory language follows the spectral-memory framing in \citet{kalathoor2026interfaceMemoryPreprint}. The present paper uses those unsheared route families as a reference state and asks what survives when the mixed route is perturbed by an imposed mean shear.

The simulations show a decoupling between finite-depth reach and spectral pathway. Reach and contact timing survive the mean shear, while the spectral branch changes. A finite-depth plume route can preserve where it reaches while changing how it remains connected.

\section{Configuration and Route Measures}

\subsection{Simulation Set}

All simulations use the same nondimensional two-layer fingering-favorable configuration with density ratio \(\Rrho=1.2\), Prandtl number \(\Pran=7\), diffusivity ratio \(\tau=0.01\), temperature jump \(1\), and interface thickness \(h_i=3\). The domain dimensions are \((L_x,L_y,L_z)=(164.367,82.184,164.367)\). The four no-shear route-atlas runs use a \(384\times192\times960\) grid and are advanced to \(t=60\) with \(\Delta t=0.0015\). The horizontal directions are periodic. The vertical direction is bounded, with far-field relaxation zones placed away from the initial interface. These zones provide a finite-depth setting: the early route atlas develops in the interior, while the zones define the remote layers that late plume forests can approach.

The no-shear atlas varies only the imposed interfacial roughness spectrum: low-mode, high-annulus, mixed, and an independent mixed-seed realization. \Cref{tab:case-nomenclature} defines the compact symbols used in figures and tables. The mean-shear run, \(M+U_s\), perturbs the mixed route \(M\). The imposed initial velocity is
\begin{equation}
u_s(z) = U_0 \tanh\!\left(\frac{z-L_z/2}{\delta_s}\right),
\qquad U_0=0.10,\qquad \delta_s=2h_i.
\label{eq:shear-profile}
\end{equation}
This shear is imposed as part of the initial condition and then evolves with the flow. The shear calculation uses the same grid, timestep, and analysis interval as the route-atlas simulations.

This design keeps the perturbation question narrow. The shear run is a single controlled initial-condition experiment. It asks whether the route selected by the mixed interface remains identifiable when the initial condition also contains a coherent, large-scale horizontal velocity profile. Because the forcing is absent after initialization, any late mean profile, spectral redistribution, or finite-depth contact must emerge from the coupled evolution of the sheared initial state and the salt-finger plume forest.

The integrations are performed with Oceananigans \citep{ramadhan2020oceananigans,silvestri2023oceananigans,wagner2025oceananigans}, using a nonhydrostatic pressure projection and centered second-order advection. The analysis uses saved mid-plane fields, mean profiles, point probes, and selected full-volume fields. A finer mixed-route run at \(512\times256\times1280\) supports the reference route at selected comparison times.

\begin{table}[tbp]
\centering
\caption{Compact case notation. The symbols are used in the figures to keep legends short while preserving the role of each simulation.}
\label{tab:case-nomenclature}
\input{tables/case_nomenclature.tex}
\end{table}

\subsection{Active Width and Contact}

For a field \(q\), let \(A_q(z,t)\) denote a vertical activity profile. For vertical velocity, \(A_w\) is the horizontal root-mean-square profile of \(w\). For salinity, \(A_S\) is the horizontal root-mean-square profile of the salinity fluctuation. Given a threshold fraction \(\alpha\), the active set is
\begin{equation}
\mathcal{A}_q(t;\alpha)
= \left\{z:\ A_q(z,t) \ge \alpha \max_z A_q(z,t)\right\}.
\label{eq:active-set}
\end{equation}
The active width is
\begin{equation}
W_q(t;\alpha) =
\max \mathcal{A}_q(t;\alpha) - \min \mathcal{A}_q(t;\alpha),
\label{eq:active-width}
\end{equation}
and contact occurs when \(\mathcal{A}_q\) reaches the inner edge of either far-field relaxation region. The corresponding contact time is
\begin{equation}
t_{c,q} = \inf\{t:\ d_q(t;\alpha) \le 0\},
\label{eq:contact-time}
\end{equation}
where \(d_q\) is the smaller distance between the active set and the two inner relaxation-zone edges. A standard threshold gives the main contact-time values, and a separate threshold-sensitivity calculation tests how strongly the finite-depth contact label depends on that choice. This keeps the endpoint route labels and the mixed-route scalar-contact sensitivity distinct.

\subsection{Spectral Branch and Transport}

Planform spectra are used to separate route memory from vertical reach. If \(\widehat{q}(\boldsymbol{k},t)\) is the horizontal Fourier transform of a mid-plane field and \(P_q(\boldsymbol{k},t)=|\widehat{q}|^2\), the fraction in a wavenumber band \(\mathcal{B}\) is
\begin{equation}
f_q(\mathcal{B},t) =
\frac{\sum_{\boldsymbol{k}\in\mathcal{B}}P_q(\boldsymbol{k},t)}
     {\sum_{\boldsymbol{k}}P_q(\boldsymbol{k},t)}.
\label{eq:spectral-fraction}
\end{equation}
We use broad, intermediate, and short-wave bands. Broad power corresponds to domain-scale connection, intermediate power carries oblique or annular route memory, and short-wave power measures retained small-scale structure.

The scalar-exchange measure is the down-gradient salinity flux,
\begin{equation}
F_S(t) = -\langle w S' \rangle,
\label{eq:salt-flux}
\end{equation}
with the sign convention chosen so positive \(F_S\) denotes down-gradient salinity transport. The cumulative exchange is
\begin{equation}
Q_S(t) = \int_0^t F_S(t')\,dt'.
\label{eq:cumulative-exchange}
\end{equation}
Flux is interpreted together with contact time and active width. A route that has large flux because it is broad and connecting is physically different from a route that retains spectral memory but remains localized.

\subsection{Route-Survival Test}

The mixed and mixed-seed no-shear runs define a route-family tolerance layer. For a quantity \(X\), we compare the sheared mixed route with the unsheared mixed reference and the independent mixed-seed realization. Over the mature post-\(t=45\) window, the mixed/seed differences provide an empirical measure of realization-to-realization variability. A sheared route that stays inside this tolerance in reach but leaves it in spectral partition has reach survival with spectral-route modification. This statement is more specific than a binary survive/fail label and avoids using scalar contact alone as the route criterion.

This definition deliberately separates three outcomes that a single scalar score blends. A perturbation can preserve reach and exchange while changing the spectral branch; it can suppress reach while leaving some local spectral memory; or it can accelerate broad connection toward the low-mode endpoint. Survival is treated as a component-wise statement. When reach, contact timing, and active width remain within the mixed-route tolerance, but spectral fractions leave that tolerance, the result is route survival with pathway change.

The five calculations are compared through this hierarchy. The first four simulations define the no-shear route space; they are not four unrelated cases ordered by a single metric. Low-mode roughness, high-annulus roughness, and mixed roughness identify the broad connecting endpoint, the localized route-memory endpoint, and the delayed transfer route. The mixed-seed realization gives the reproducibility scale of that delayed route. The mean-shear calculation is then an intervention on the mixed route. It tests whether imposed shear destroys the delayed route, redirects it toward either no-shear endpoint, or preserves its finite-depth reach while altering the route pathway.

\section{Results and Discussion}

The results are organized around the route hierarchy. The no-shear calculations first define the route atlas: \cref{tab:route-components,fig:route-connection-time} identify the connecting, localized-memory, and delayed-transfer routes before any shear is imposed. Next, \cref{fig:contact-width,fig:distance-to-layer,fig:shear-phase-space} place the mean-shear intervention \(M+U_s\) back into that atlas using directly comparable reach, contact, and spectral coordinates. Finally, \cref{fig:mixed-seed-ratios,fig:threshold-sensitivity,tab:route-transport} show which mixed-route measures are reproducible before shear is added, while \cref{fig:shear-reach,fig:shear-time-resolved,fig:shear-spectral-profile,tab:final-shear,tab:selected-3d} test the shear intervention against that tolerance layer.

\subsection{No-Shear Atlas: Multiple Route Components}

The no-shear atlas establishes the comparison space before mean shear is introduced. The four reference calculations separate along several axes: route connection, contact, vertical extent, active area, spectral memory, and probe activity (\cref{tab:route-components}). The low-mode route has the largest route-connection index, \(\Rc=0.9625\), with contact, vertical extent, and active-area scores all at or near the top of the atlas. It is the broad connecting endpoint: the active layer reaches the finite-depth boundaries earlier, spreads over a large vertical interval, and carries the largest exchange.

The high-annulus route has a different component balance. No distance-based velocity or salinity contact occurs by \(t=60\), and its final route-connection index is \(0.2356\). At the same time, it has the largest spectral-memory score, \(0.8456\), exceeding the low-mode value \(0.7497\). That combination matters physically. The high-annulus case retains a coherent planform imprint of the imposed annular roughness while the active layer remains localized relative to the broad low-mode route. Its final flux and cumulative exchange, \(0.09812\) and \(1.797\), are weak compared with the broad connecting endpoint, even though the planform structure remains organized.

The high-annulus calculation has its own route identity. Its spectral organization is strong, and that organization is concentrated in a branch that does not efficiently build finite-depth reach over the analyzed window. A route-memory endpoint can be coherent in planform and weak in exchange at the same time. That coexistence motivates keeping route memory, contact, and flux as separate quantities.

The mixed route occupies a delayed-transfer state between these endpoints. It stays closer to the localized side through the middle of the run and then approaches finite-depth connection late. Its distance-based velocity contact occurs at \(t=57.75\), salinity contact occurs at \(t=59.5\), and its final low--high route coordinate is \(-0.1145\), much closer to the localized side than to the low-mode endpoint even after contact has occurred. Its final salt flux \(F_S=0.1595\) and cumulative exchange \(Q_S=2.600\) are intermediate. The mixed route is a delayed pathway between the endpoint spectra: initially short or intermediate scales feed a broader finite-depth plume forest only late in the integration.

The mixed route also supplies the most useful perturbation target. It is close enough to finite-depth contact that a perturbation can plausibly retime or reshape connection, while still remaining distinct from the early broad connecting state. It also retains enough spectral structure that pathway changes can be measured. Reach, memory, and transport are all active enough in the mixed case to test whether a plume route survives added large-scale motion.

\begin{figure}[tbp]
\centering
\begin{subfigure}{0.49\linewidth}
\centering
\includegraphics[width=\linewidth]{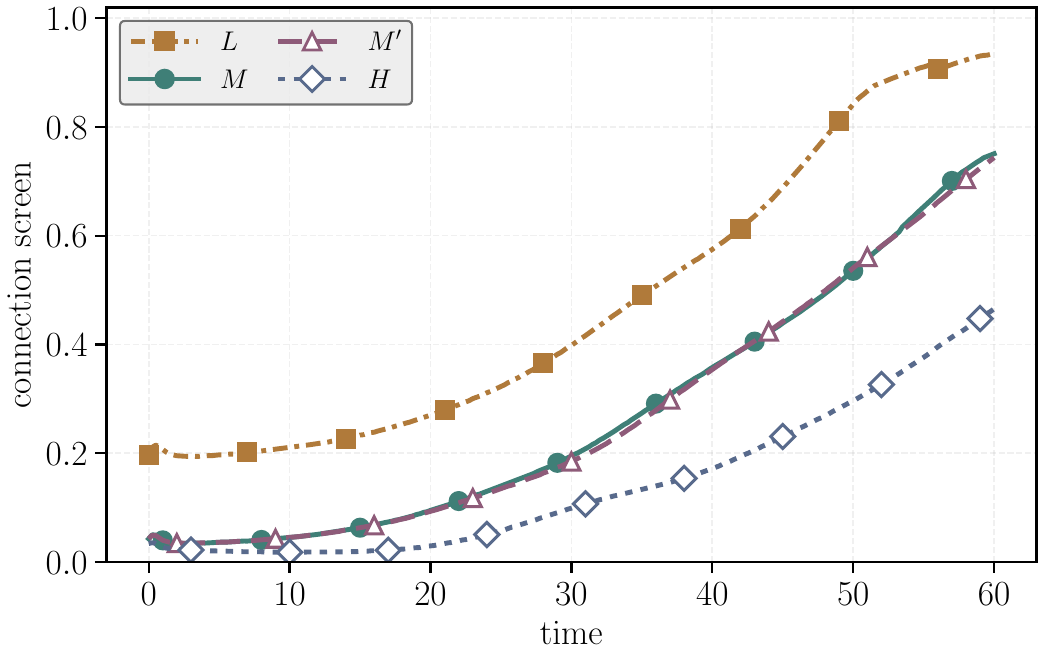}
\caption{Connection index}
\label{fig:route-connection-index}
\end{subfigure}
\hfill
\begin{subfigure}{0.49\linewidth}
\centering
\includegraphics[width=\linewidth]{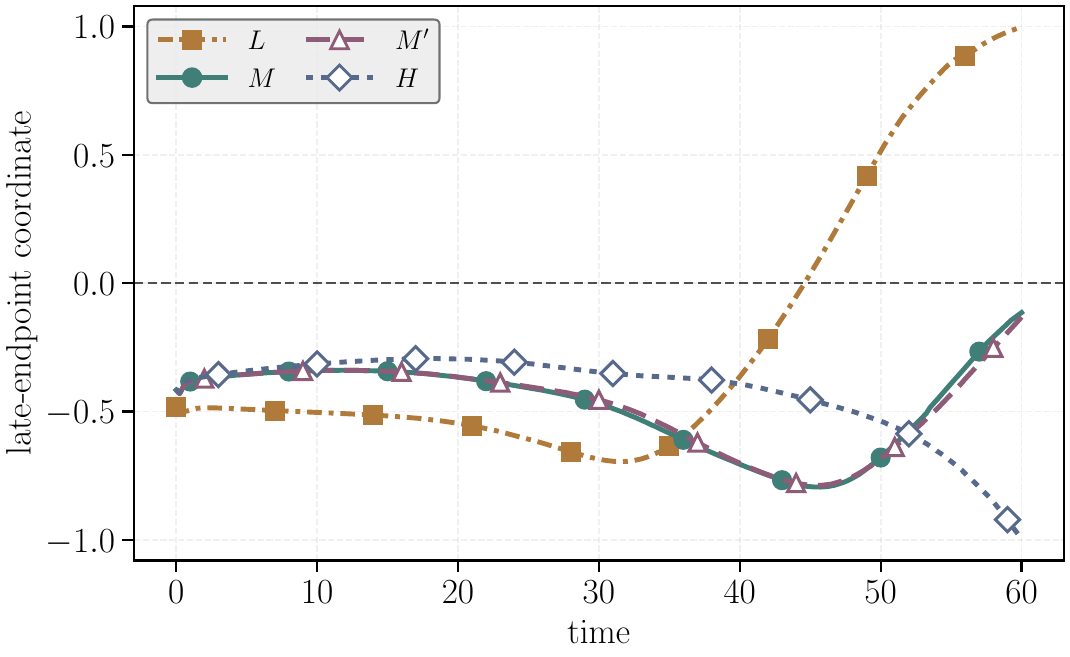}
\caption{Low--high coordinate}
\label{fig:route-coordinate}
\end{subfigure}
\caption{No-shear route coordinates. \(L\) reaches the connecting state first, \(M\) approaches connection late, and \(H\) remains localized through the simulated interval.}
\label{fig:route-connection-time}
\end{figure}

\Cref{fig:route-connection-time} shows how this separation develops in time. Low-mode roughness reaches \(\Rc=0.5\) at \(t=35.75\) and \(\Rc=0.7\) at \(t=45.5\). The mixed route reaches the same thresholds much later, at \(t=48.5\) and \(t=57.0\). The high-annulus route reaches neither threshold by \(t=60\). The low--high route coordinate gives the same ordering at the end of the run: low mode is at \(1.0\), high annulus is at \(-1.0\), mixed is at \(-0.1145\), and the mixed-seed realization is at \(-0.1322\). The finite-depth implication is that \(L\) is the early connecting endpoint, \(H\) is the localized endpoint, and \(M/M'\) approach connection late. The direct reach histories are used below to place the mean-shear intervention in the same coordinates.

\begin{table}[tbp]
\centering
\caption{Route components for the no-shear atlas. \(R_c\) is the route-connection index and \(C_p\) is the plume-coherence index. The component scores show that route coherence has several separable parts.}
\label{tab:route-components}
\input{tables/route_components.tex}
\end{table}

\subsection{Mixed-Route Tolerance: Continuous Similarity and Scalar-Contact Sensitivity}

\begin{figure}[tbp]
\centering
\includegraphics[width=0.82\linewidth]{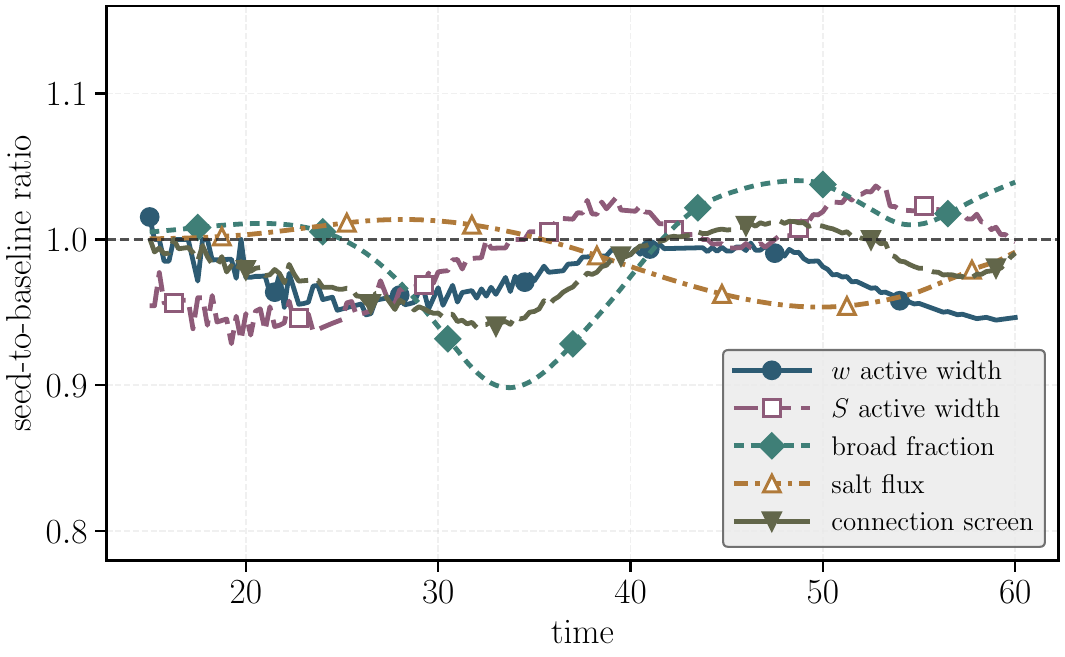}
\caption{Time-resolved mixed-seed ratios relative to the mixed baseline. Continuous route metrics remain close after the plume forest has developed, which defines the tolerance layer used for the shear perturbation.}
\label{fig:mixed-seed-ratios}
\end{figure}

The mixed-seed replicate defines the tolerance of the delayed route family. A turbulent plume forest changes pointwise under a new seed or a small perturbation, so persistence is judged at the route level: late active width, spectral partition, exchange, and approximate timing relative to the endpoint routes. \Cref{fig:mixed-seed-ratios} shows that this route-level behavior is reproducible after the plume forest has developed. Over the post-\(t=45\) window, the mean absolute relative differences between the mixed and mixed-seed realizations are \(3.1\%\) for \(w\)-active width, \(1.6\%\) for salinity-active width, \(2.8\%\) for broad spectral fraction, \(4.2\%\) for intermediate spectral fraction, \(3.6\%\) for salt flux, and \(2.4\%\) for cumulative exchange. The final fluxes, \(0.1595\) and \(0.1582\), and the final cumulative exchanges, \(2.600\) and \(2.530\), are close on the scale of the endpoint separation.

\begin{figure}[tbp]
\centering
\begin{subfigure}{0.49\linewidth}
\centering
\includegraphics[width=\linewidth]{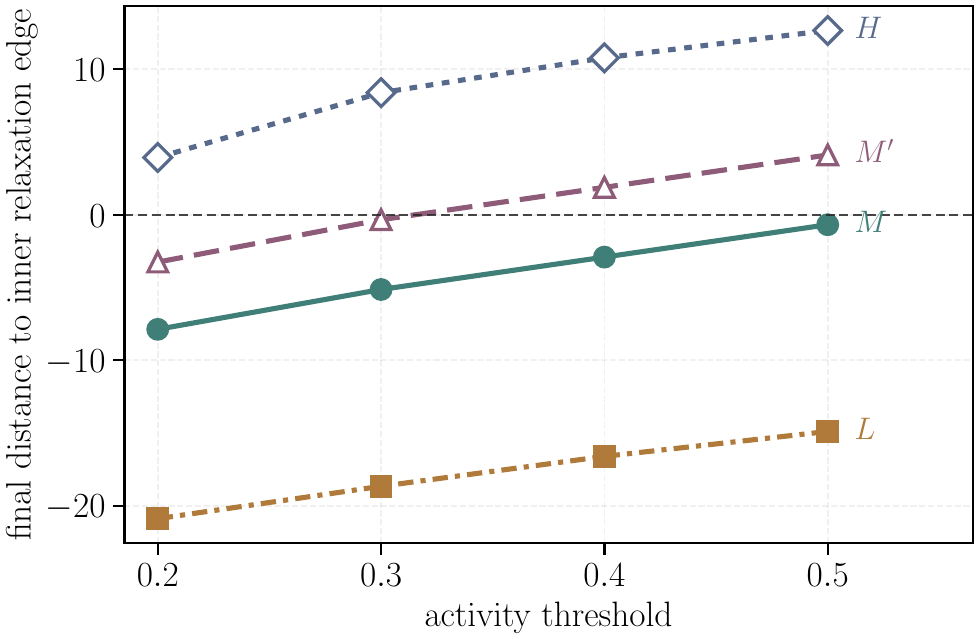}
\caption{\(w\)}
\label{fig:threshold-w}
\end{subfigure}
\hfill
\begin{subfigure}{0.49\linewidth}
\centering
\includegraphics[width=\linewidth]{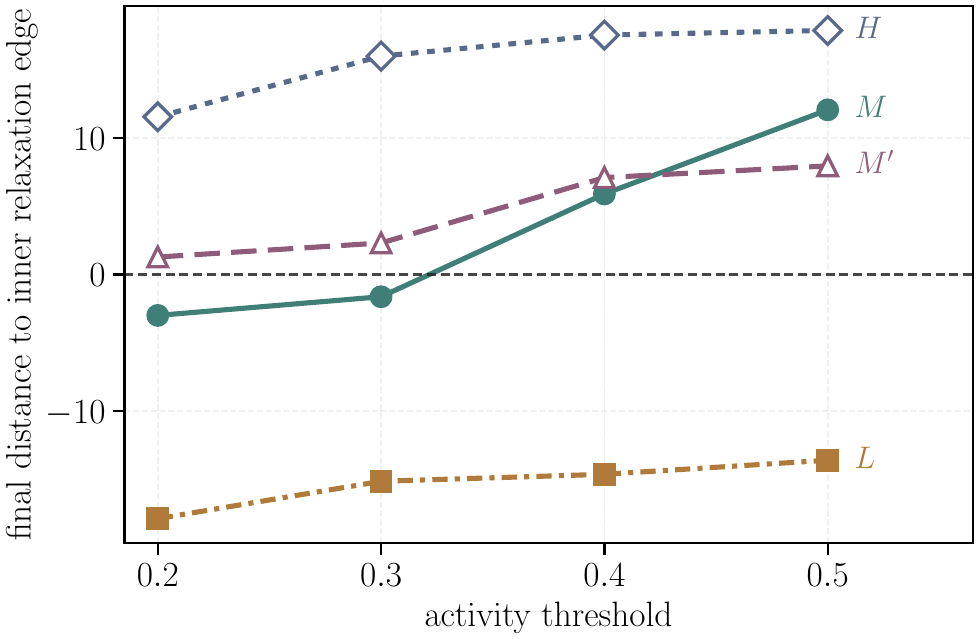}
\caption{Salinity}
\label{fig:threshold-s}
\end{subfigure}
\caption{Threshold sensitivity of final distance to the relaxation region. Endpoint contact outcomes are robust, while mixed-route scalar contact is threshold-sensitive.}
\label{fig:threshold-sensitivity}
\end{figure}

The same replicate identifies the threshold-sensitive part of the mixed route. Binary scalar contact near \(t=60\) changes with the activity threshold. Across the tested threshold set, the mixed realization gives salinity contact in \(2/4\) thresholds, whereas the mixed-seed realization gives salinity contact in \(0/4\). Velocity contact is more persistent, with \(4/4\) thresholds in the mixed realization and \(2/4\) in the mixed-seed realization. The endpoint routes are stable under the same test: low-mode roughness contacts in both fields for all thresholds, while high-annulus roughness avoids contact in both fields for all thresholds (\cref{fig:threshold-sensitivity}).

This behavior motivates the combined use of continuous and thresholded route measures. The mixed route lies near a scalar-contact boundary; a small shift in salinity envelope can change a binary late-contact label while leaving the underlying route family nearly unchanged. The shear test below is judged by active width, contact timing, spectral partition, and exchange together. A shear case that preserves reach while altering spectral partition is physically different from a case whose main effect is a small delay in scalar contact.

The scalar-contact sensitivity also fixes how strong the shear claim can be. A salinity-contact flag alone separates the mixed and mixed-seed realizations into different classes at \(t=60\). The continuous measures show a different picture: active widths, spectral fractions, flux, and cumulative exchange remain close after \(t=45\). The route-family comparison is based on the quantities that are stable under a seed change, while the thresholded contact flags identify the narrow part of the response that is intrinsically sensitive near the end of the run.

\subsection{Exchange Follows Route Geometry}

The route atlas has a transport consequence in addition to its visual and geometric separation. \Cref{tab:route-transport} shows that final salt flux and cumulative exchange follow the route ordering. The broad low-mode endpoint has final flux \(0.3009\) and cumulative exchange \(6.438\). The localized high-annulus endpoint has final flux \(0.09812\) and cumulative exchange \(1.797\). The two mixed realizations sit between them, with final fluxes \(0.1595\) and \(0.1582\), and cumulative exchanges \(2.600\) and \(2.530\).

These differences are large enough to matter for the physical interpretation. The low-mode route carries about \(3.07\) times the final flux and \(3.58\) times the cumulative exchange of the high-annulus route. Relative to the mixed route, the low-mode endpoint carries about \(1.89\) times the final flux and \(2.48\) times the cumulative exchange. Thus the broad connecting endpoint is a higher-exchange finite-depth state as well as a different morphology. Conversely, high spectral memory without finite-depth reach is a lower-flux state. The mixed route gives intermediate exchange because broad connection develops only late in the run.

This transport separation clarifies what the route measures add. Final-flux ordering identifies high-annulus roughness as weak and mixed roughness as intermediate, but it does not explain the route by which those fluxes are obtained. High-annulus roughness preserves planform memory while remaining localized away from the remote layers. Mixed roughness transfers gradually into broader activity and reaches the relaxation-region vicinity late. The transport difference follows from both plume strength and the vertical path taken by the active plume forest.

Cumulative exchange strengthens this reading because it records when connection occurred, in addition to how large the final flux became. Low-mode roughness has both early route connection and the largest final flux, so it accumulates much more exchange over the full interval. Mixed roughness reaches connection late; its final flux is intermediate, but its cumulative exchange also carries the cost of delayed broadening. High-annulus roughness maintains organized spectral memory without comparable reach, giving the weakest cumulative exchange. Thus the transport result carries time-history information: it ties the active-layer growth path to the amount of scalar exchange produced by the route.

\begin{table}[tbp]
\centering
\caption{Transport and route-connection summary for the no-shear atlas.}
\label{tab:route-transport}
\input{tables/route_transport.tex}
\end{table}

\subsection{Mean Shear as a Route-Survival Test}

The shear perturbation tests the delayed mixed route against the no-shear route atlas. This route is the natural target because it is late enough for contact timing to be shifted, close enough to the scalar-contact boundary for finite-depth reach to be modified, and reproducible enough to define a mixed/seed tolerance layer. The imposed initial profile \cref{eq:shear-profile} gives a direct route-survival test without defining a new endpoint.

The production shear calculation applies a single finite-amplitude initial profile with \(U_0=0.10\). The amplitude is used as a controlled perturbation of the mixed route, and the response is evaluated against the no-shear route atlas and the mixed/seed tolerance layer. The shear is imposed at initialization and then evolves freely. The imposed profile must persist, decay, or be reshaped by the plume forest itself. The late response is the coupled evolution of an initial large-scale shear and a finite-depth salt-finger plume forest.

This setup separates suppression, acceleration, and route survival with a changed pathway. The full-resolution run shows the third response. Mean shear preserves the mixed-route plume forest's finite-depth reach, because the sheared active layer reaches the remote layers at the mixed-route contact times. The active widths and delayed contact remain close to \(M\), not to the earlier, broader \(L\) endpoint, and the sheared run contacts the remote layers while \(H\) remains localized through \(t=60\). The quantity that changes most clearly is the spectral pathway that accompanies the preserved reach.

\subsection{Shear Placement in the Route Atlas}

\begin{figure}[tbp]
\centering
\includegraphics[width=0.82\linewidth]{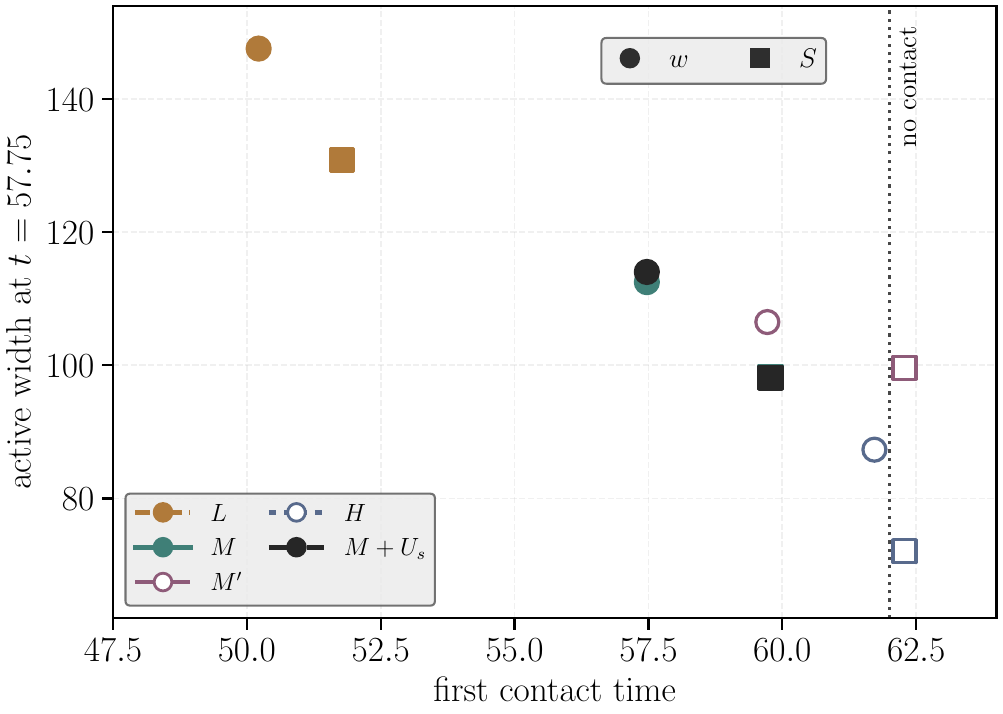}
\caption{Contact time and late active width for the no-shear route atlas and the mean-shear intervention. Markers are colored by case and shaped by field. \(M+U_s\) lies close to \(M\) in contact timing and active width, while \(L\) and \(H\) mark the broad-connecting and localized endpoints.}
\label{fig:contact-width}
\end{figure}

\Cref{fig:contact-width} places the mean-shear run in the same contact--width coordinates as the no-shear atlas. The endpoint separation is large. \(L\) combines early first contact (\(t=50.5\) for \(w\), \(t=51.5\) for salinity) with broad late active widths (\(147.6\) and \(130.8\)). \(H\) has no contact through \(t=60\); its plotted no-contact markers are placed beyond the simulated window, and its late widths, \(87.32\) for \(w\) and \(72.08\) for salinity, remain much smaller. \(M\) lies between them: first contact occurs at \(t=57.75\) for \(w\) and \(t=59.5\) for salinity, with active widths \(112.5\) and \(98.28\).

The sheared run \(M+U_s\) falls on the mixed side of this atlas, away from both endpoints. In the same contact--width plane, \(M+U_s\) has first \(w\) contact at \(t=57.75\) and first salinity contact at \(t=59.5\), with plotted active widths \(114.0\) and \(98.1\). These values are close to the mixed route and far from the early broad \(L\) endpoint. They also differ from \(H\), which retains spectral memory but never contacts the finite-depth relaxation regions over the simulated interval.

\begin{figure}[tbp]
\centering
\begin{subfigure}{0.49\linewidth}
\centering
\includegraphics[width=\linewidth]{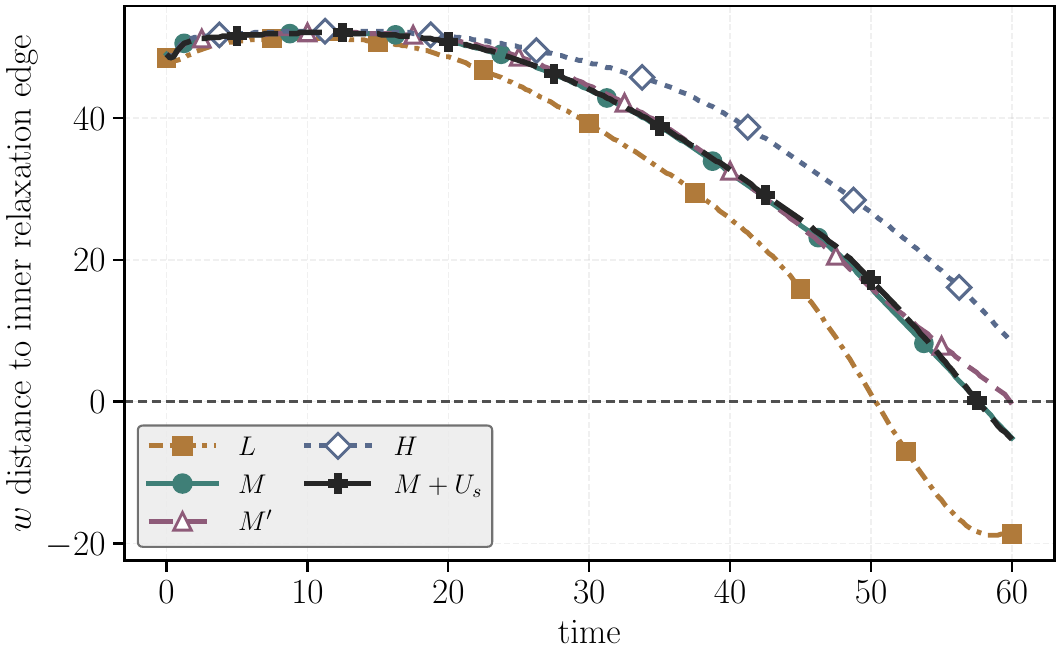}
\caption{\(w\)}
\label{fig:distance-w}
\end{subfigure}
\hfill
\begin{subfigure}{0.49\linewidth}
\centering
\includegraphics[width=\linewidth]{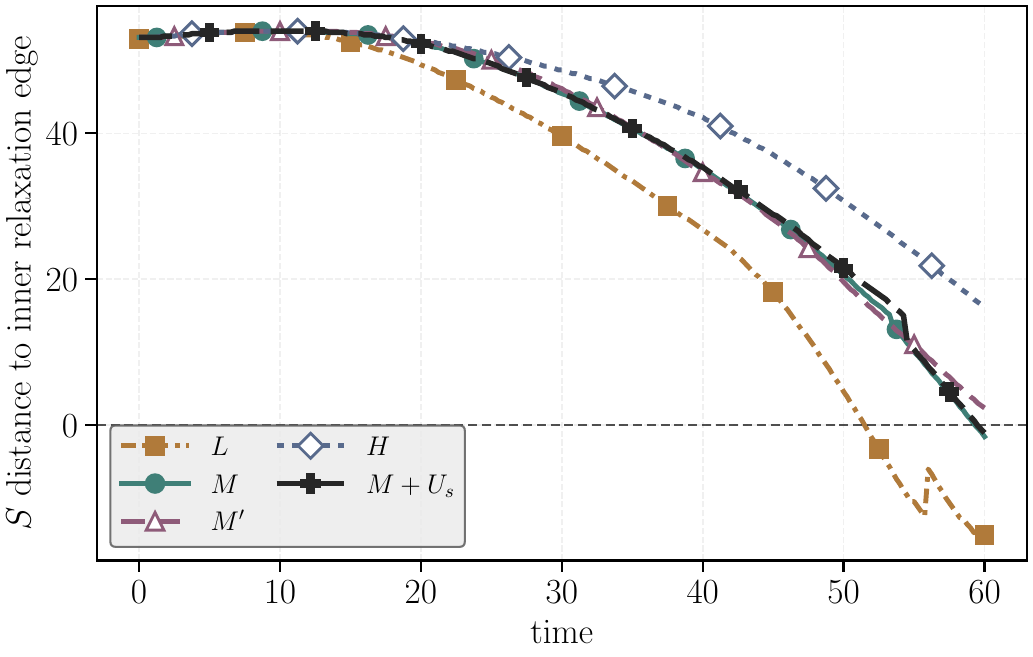}
\caption{Salinity}
\label{fig:distance-s}
\end{subfigure}
\caption{Distance from the active layer to the inner edge of the vertical relaxation regions for the no-shear atlas and the mean-shear intervention. Zero marks finite-depth contact.}
\label{fig:distance-to-layer}
\end{figure}

\Cref{fig:distance-to-layer} gives the time-resolved version of the same placement. \(L\) approaches the finite-depth layers earliest, \(H\) remains separated, and \(M/M'\) approach the inner relaxation edge late. The sheared run follows the late mixed-route approach. It does not accelerate into the early broad \(L\) route, and it does not remain localized like \(H\). These reach coordinates establish the geometric part of route survival before the spectral branch is examined.

\subsection{Reach Survival and Spectral-Pathway Change}

\begin{figure}[tbp]
\centering
\begin{subfigure}{0.49\linewidth}
\centering
\includegraphics[width=\linewidth]{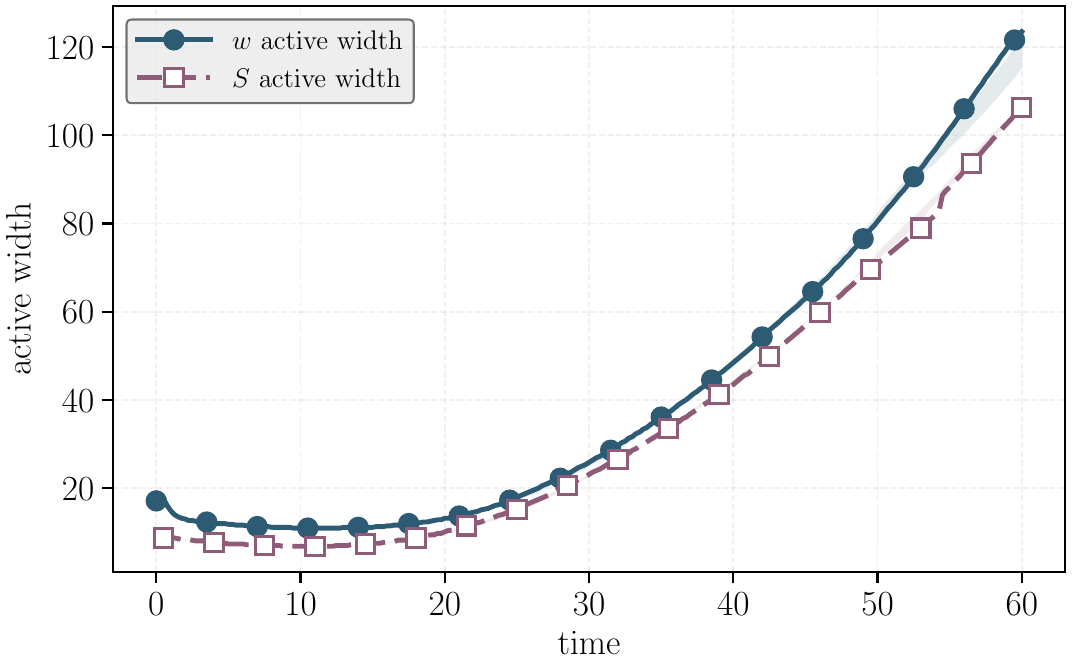}
\caption{Active width}
\label{fig:shear-active-width}
\end{subfigure}
\hfill
\begin{subfigure}{0.49\linewidth}
\centering
\includegraphics[width=\linewidth]{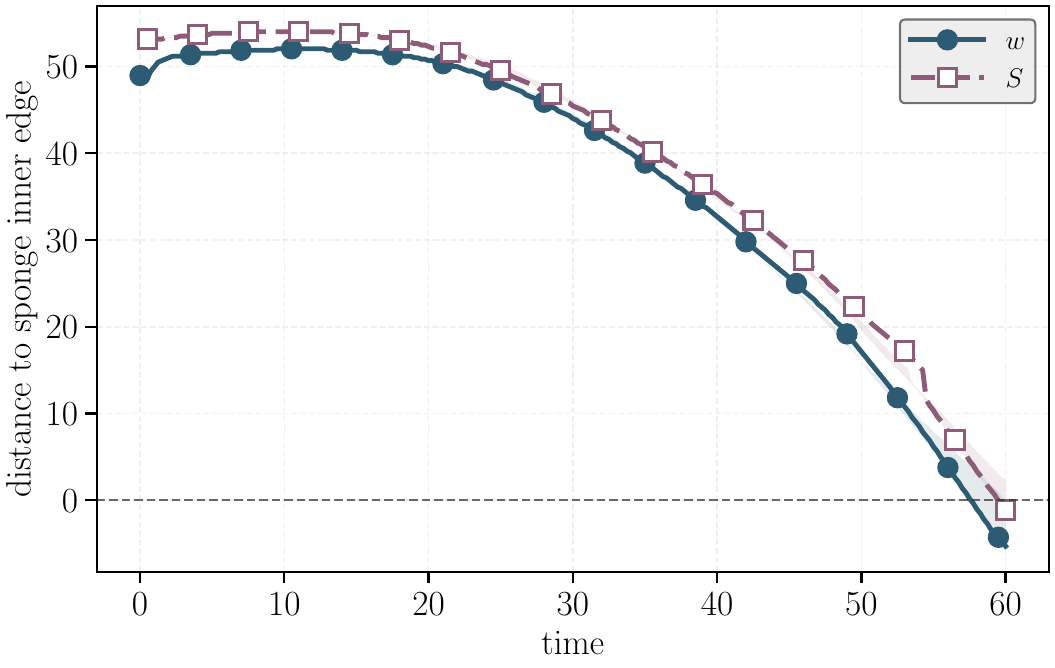}
\caption{Finite-depth reach}
\label{fig:shear-vertical-reach}
\end{subfigure}
\caption{Full-resolution shear response compared with the mixed/seed tolerance envelope. The shear case preserves the finite-depth reach of the mixed route.}
\label{fig:shear-reach}
\end{figure}

\begin{figure}[tbp]
\centering
\begin{subfigure}{0.49\linewidth}
\centering
\includegraphics[width=\linewidth]{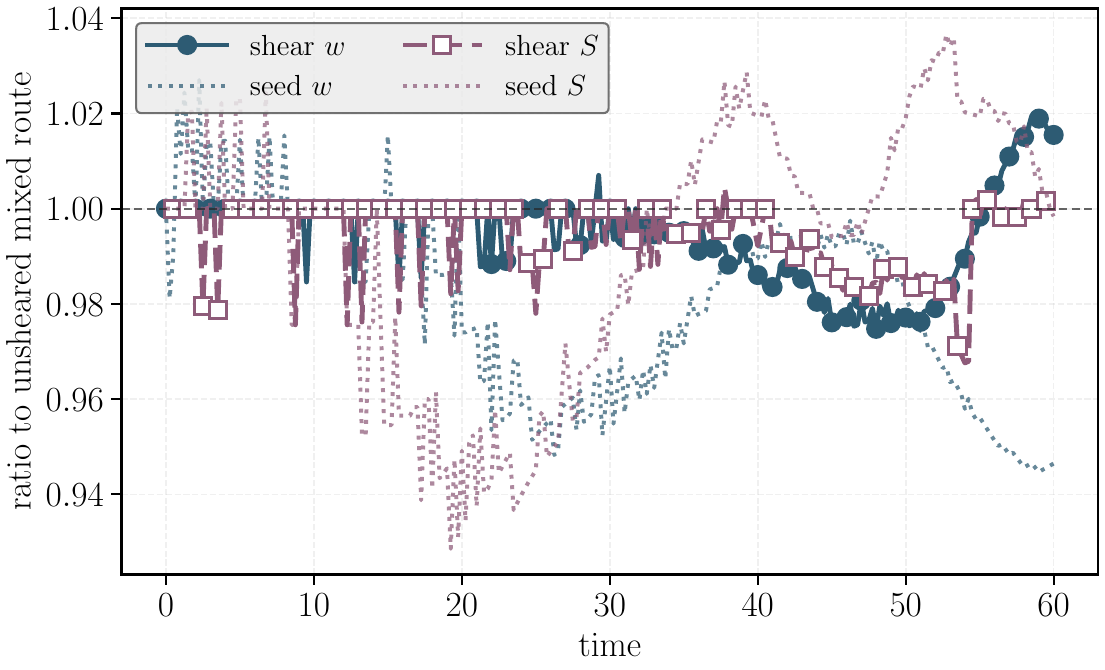}
\caption{Active-width ratios}
\label{fig:shear-width-ratios}
\end{subfigure}
\hfill
\begin{subfigure}{0.49\linewidth}
\centering
\includegraphics[width=\linewidth]{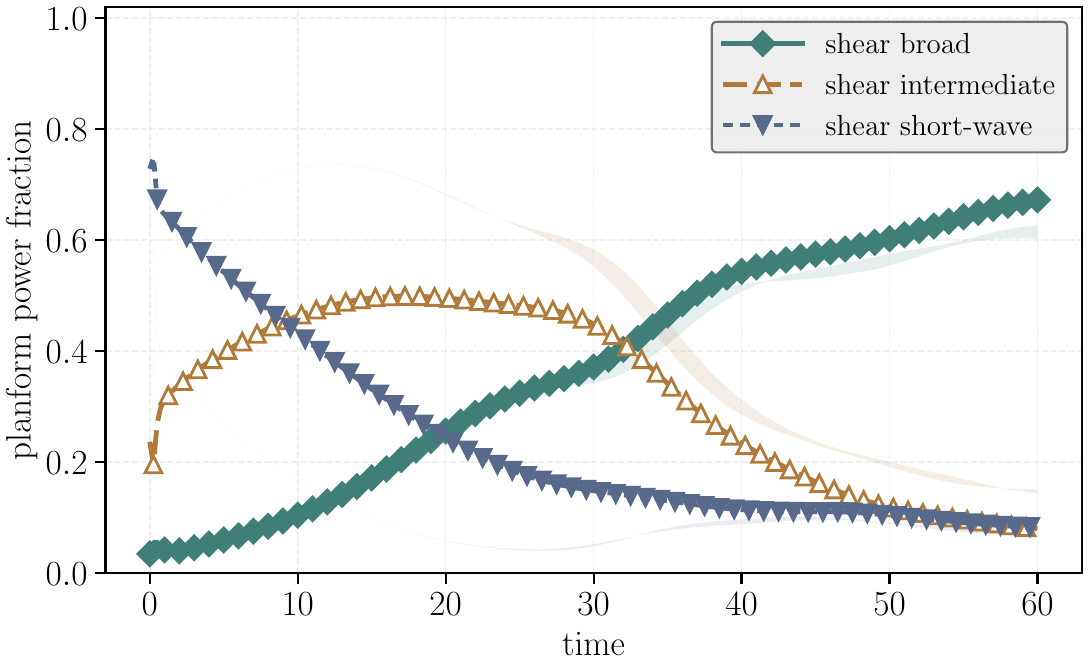}
\caption{Spectral fractions}
\label{fig:shear-spectral-ratios}
\end{subfigure}
\caption{Time-resolved full-resolution shear response relative to the unsheared mixed route and mixed-seed reference. Active widths stay close to the mixed route after the plume forest has developed, while the spectral partition remains shifted throughout the late interval.}
\label{fig:shear-time-resolved}
\end{figure}

The full-resolution \(U_0=0.10\) run gives the central route-survival result. \Cref{fig:shear-time-resolved,fig:shear-reach} and \cref{tab:final-shear} show that the sheared run preserves the finite-depth reach of the mixed route while changing the spectral branch. First velocity contact is at \(t=57.75\), exactly matching the unsheared mixed route. First salinity contact is at \(t=59.5\), also matching mixed. The final active widths remain inside the mixed/seed tolerance layer: \(w\)-active width is \(123.447\) in the sheared run, compared with \(121.563\) in mixed and \(115.057\) in mixed-seed; salinity-active width is \(106.325\), compared with \(106.154\) and \(105.983\). At \(t=60\), the sheared active widths differ from mixed by only \(1.5\%\) for \(w\) and \(0.2\%\) for salinity.

These numbers identify the sheared response as route survival of the delayed mixed family. The sheared plume forest retains late contact and broadens to nearly the same active widths as the unsheared mixed route. Its contact times also match the mixed route instead of moving toward the earlier low-mode contacts. The preserved quantity is finite-depth reach of the delayed route, with no wholesale transfer to either endpoint route.

The time histories show that this is not only a final-time coincidence. Over \(45\le t\le60\), the sheared/mixed active-width ratios remain close to unity: \(0.975\)--\(1.019\) for \(w\) and \(0.968\)--\(1.002\) for salinity. Over the same interval, the spectral ratios separate persistently from the mixed route: the broad fraction remains above the mixed value (\(1.079\)--\(1.116\)), the intermediate fraction remains depleted (\(0.530\)--\(0.712\)), and the short-wave fraction remains elevated (\(1.155\)--\(1.278\)). The late response is organized around a specific decoupling. The plume forest keeps the mixed route's vertical reach and contact timing while using a different distribution of planform scales.

\begin{figure}[tbp]
\centering
\includegraphics[width=0.82\linewidth]{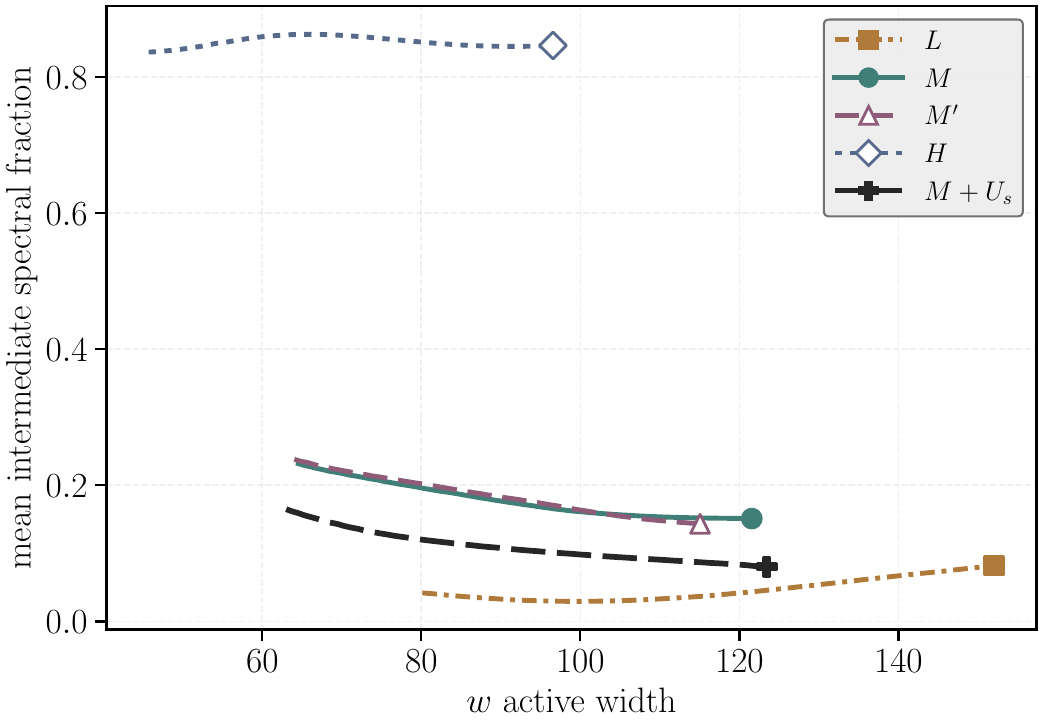}
\caption{Late-time route phase space for the no-shear atlas and the imposed shear intervention, including all five calculations. The sheared run remains close to the mixed route in active width while shifting to a lower intermediate spectral fraction.}
\label{fig:shear-phase-space}
\end{figure}

\Cref{fig:shear-phase-space} places this result back into the full route atlas. The figure plots late-time \(w\)-active width against the intermediate spectral fraction for all no-shear routes and for the shear intervention. In late-window mean values, the sheared run has nearly the same active width as the mixed route: \(91.59\) compared with \(92.18\), a shear/mixed ratio of \(0.991\). Its intermediate spectral fraction is much lower, \(0.112\) compared with \(0.180\), giving a shear/mixed ratio of \(0.618\). The final values show the same separation: the sheared \(w\)-active width is \(123.45\), close to the mixed value \(121.56\), while the intermediate fraction is \(0.080\), below the mixed value \(0.151\) and far below the high-annulus value \(0.846\).

The phase-space view resolves the route question in one plane. The shear trajectory stays away from the localized high-annulus state because its active width remains in the mixed-route range and it reaches the finite-depth layers. It also remains distinct from the low-mode endpoint, which is broader and contacts earlier. The sheared run occupies a mixed-route reach coordinate with a shifted spectral coordinate. This is the clearest graphical statement of route survival with spectral modification.

The spectral branch changes in the same run. At \(t=60\), the broad fraction is \(1.116\) times the mixed value, the intermediate fraction is \(0.530\) times the mixed value, and the short-wave fraction is \(1.278\) times the mixed value. Over \(45\le t\le60\), the ratios remain separated from the mixed/seed tolerance layer: \(1.088\) for broad fraction, \(0.618\) for intermediate fraction, and \(1.202\) for short-wave fraction. The direction of the change is physically informative. The sheared run still develops broad connection, but it carries less intermediate-scale memory and more short-wave content than the unsheared mixed route. Shear leaves finite-depth reach intact while changing the spectral pathway by which that reach is maintained.

The spectral change is large compared with the mixed/seed tolerance layer. Active widths remain within a few percent of the unsheared mixed route, but the intermediate fraction is nearly halved and the short-wave fraction is elevated. The sheared run reaches the same finite-depth layers through a different flow organization. It develops broad connection while retaining more small-scale structure and less intermediate-scale memory than the unsheared mixed route. This is the main physical distinction between route survival with pathway change and an unchanged replay of the unsheared route.

\begin{figure}[tbp]
\centering
\begin{subfigure}{0.49\linewidth}
\centering
\includegraphics[width=\linewidth]{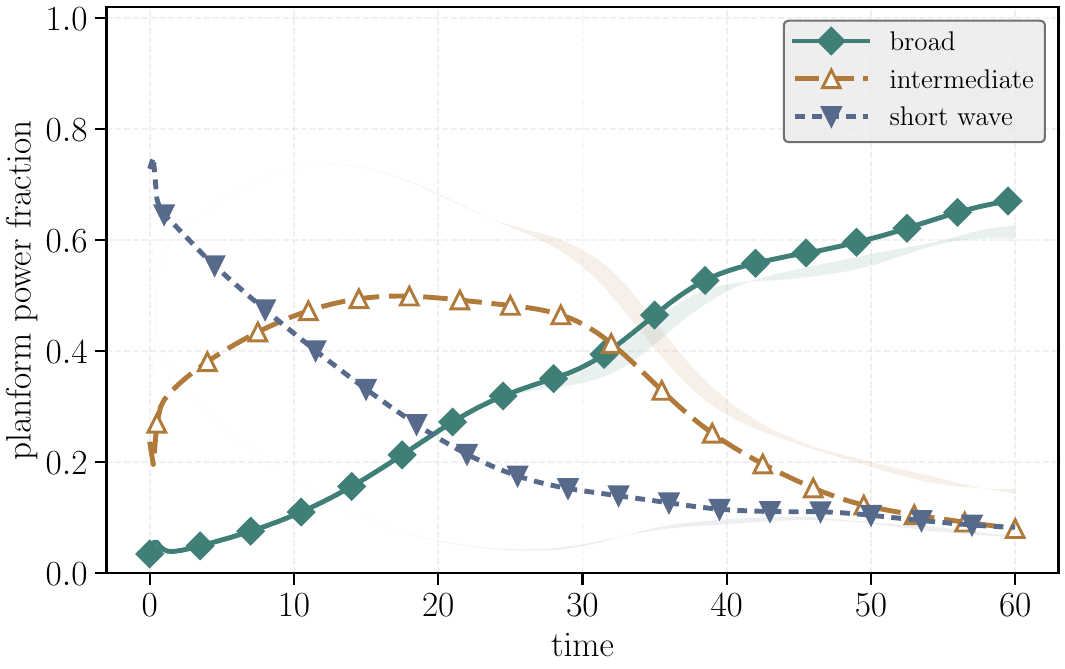}
\caption{Spectral branch}
\label{fig:shear-spectral}
\end{subfigure}
\hfill
\begin{subfigure}{0.49\linewidth}
\centering
\includegraphics[width=\linewidth]{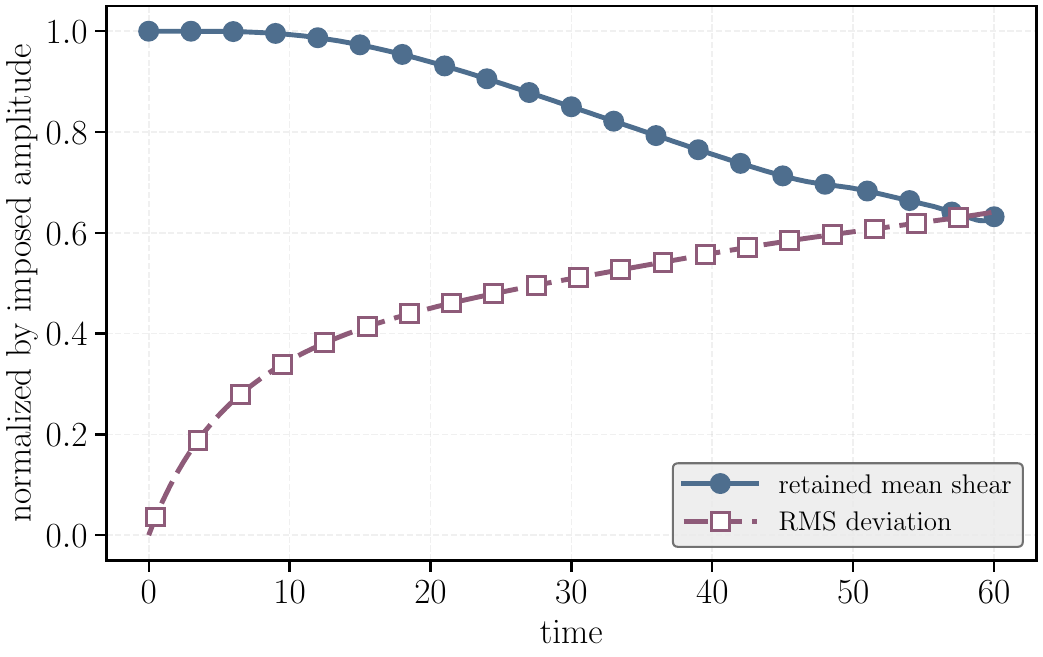}
\caption{Mean-shear profile}
\label{fig:shear-profile}
\end{subfigure}
\caption{Spectral redistribution and mean-profile retention in the sheared run. The mean profile retains \(63.2\%\) of the imposed amplitude at \(t=60\), while the RMS departure from the initial profile reaches \(64.1\%\) of the imposed amplitude.}
\label{fig:shear-spectral-profile}
\end{figure}

\begin{table}[tbp]
\centering
\caption{Final shear response against the mixed reference and the mixed-seed tolerance layer.}
\label{tab:final-shear}
\input{tables/final_shear_response.tex}
\end{table}

The mean-profile evolution links the spectral change to the coupled sheared flow. \Cref{fig:shear-spectral-profile} shows that the mean profile remains present but is strongly reshaped. At \(t=60\), the retained mean-shear amplitude is \(0.631854\) of the initial imposed amplitude, while the residual RMS relative to the imposed amplitude is \(0.641263\). The horizontally averaged flow still contains a coherent large-scale shear component, but the plume-induced departure is comparable to the retained profile. The plume forest actively reshapes the mean flow while retaining finite-depth reach and redistributing spectral content.

The mean-profile result links the spectral redistribution to the evolving large-scale flow. The imposed profile neither vanishes rapidly nor remains nearly unchanged. The measured retention and residual RMS show that a coherent mean shear component persists while the plume forest produces a comparably large departure from the initial profile. The changed spectral pathway is part of a two-way adjustment between plume activity and mean flow.

Selected full-volume checks support this response beyond the mid-plane slices. The high-cadence slice proxy shows steadily strengthening late exchange, with the salt-flux proxy reaching \(0.100197\) and mean \(|w|\) reaching \(0.529167\) at \(t=60\). The selected full-volume salt-flux proxy at the same time is \(0.116285\), so the late transport signal is not confined to a single plane. The 3D \(w\)-active width is \(131.836\), and the 3D salinity active width is \(108.893\) (\cref{tab:selected-3d}). The 3D velocity active set has already crossed the inner relaxation-zone edge by \(t=58\), with minimum distance \(-8.989\). The salinity active set is still \(2.996\) units inside the edge at \(t=58\) and reaches it by \(t=60\), with minimum distance \(-0.770\). The 3D fields reproduce the same late sequence as the planar measures: velocity reaches first, salinity follows near the end, and the finite-depth reach conclusion is reproduced beyond a single slice.

The selected volumes also clarify the ordering of field response. Vertical velocity reaches the relaxation-zone vicinity before salinity, both in the planar contact histories and in the 3D active sets. That ordering is consistent with a route in which vertical motion expands first and scalar structure follows as the plume forest fills the finite-depth layer. The shear result is not only a geometric width result; it also preserves the field-asymmetric timing already present in the mixed-route family.

\begin{table}[tbp]
\centering
\caption{Selected full-volume support checks for the full-resolution shear run.}
\label{tab:selected-3d}
\input{tables/selected_3d_checks.tex}
\end{table}

\subsection{Resolution Support and Scope}

The shear comparison requires a mixed-route reference whose late route features are not concentrated near the grid cutoff. A finer \(512\times256\times1280\) mixed run supports that reference by showing that late high-wavenumber tails are no larger in the finer run than in the \(384\times192\times960\) run at the selected comparison times (\cref{tab:resolution-support}). For salinity at \(t=60\), the z-mid tail above \(0.8 k_\mathrm{Ny}\) is \(4.81\times10^{-4}\) on the production grid and \(3.89\times10^{-5}\) on the finer grid; the corresponding \(k_{95}/k_\mathrm{Ny}\) values are \(0.2329\) and \(0.1735\). For \(w\), the z-mid tails are \(5.99\times10^{-8}\) and \(1.62\times10^{-9}\). The temperature and vertical-spectrum tails are still smaller. These comparisons support using the mixed reference route as a resolved-scale shear-test baseline.

\begin{table}[tbp]
\centering
\caption{Selected ultra-resolution support for the mixed reference route. The table reports late high-wavenumber tail measures from the production-grid and finer mixed runs.}
\label{tab:resolution-support}
\input{tables/resolution_support.tex}
\end{table}

The result is scoped to the simulations performed here: \(\Rrho=1.2\), \(\Pran=7\), \(\tau=0.01\), \(h_i=3\), fixed roughness amplitude, fixed finite-depth geometry, and one imposed initial shear design. Within that setting, the route response is precise. The imposed mean shear preserves the mixed route's finite-depth reach and late contact timing while modifying the spectral partition that maintains that reach. Flux alone misses the branch change, contact alone misses the spectral redistribution, and visual morphology alone does not quantify the mixed-route tolerance layer.

The scope is deliberately finite. The present calculations address one well-resolved configuration with fixed density ratio, interface thickness, shear width, and forcing history. Within that setting, the route selected by a mixed interface can survive in reach while changing its spectral pathway under an added large-scale velocity structure. That result motivates route-based comparisons in future finite-depth double-diffusive studies, especially when interfaces carry inherited roughness or are embedded in larger-scale motion.

\section{Conclusions}

Finite-depth salt-finger plume forests in this configuration organize into route families. Low-mode roughness produces a broad connecting route, high-annulus roughness produces a localized route-memory endpoint, and mixed roughness produces a delayed scale-transfer route. A second mixed realization shows that the delayed route is reproducible in continuous active-width, spectral, and exchange measures after \(t=45\), even though binary scalar contact near \(t=60\) is threshold-sensitive.

An imposed initial mean shear with \(U_0=0.10\) preserves the mixed route's finite-depth reach and late contact timing. The same perturbation changes the spectral pathway: broad connection is slightly enhanced, intermediate-scale memory is reduced, and short-wave content remains elevated relative to the unsheared mixed route. The mean profile also remains coherent while being reshaped by plume feedback. Route survival in this setting means preserved reach with a modified pathway relative to the unsheared route.

The broader implication is that finite-depth double-diffusive plume forests are best compared through reach, spectral pathway, and transport together when the initial interface carries finite-amplitude spectral structure or when large-scale motion perturbs the flow. These quantities can respond differently to the same perturbation, and their separation is the physical content of the route-survival result.

\section*{Statements and Declarations}

\noindent\textbf{Funding.} This research received no specific grant from any funding agency in the public, commercial, or not-for-profit sectors.

\noindent\textbf{Competing interests.} The author declares no competing interests.

\noindent\textbf{Author contributions.} Sriram P. Kalathoor performed the conceptualization, methodology, software development, data curation, analysis, visualization, and writing.

\noindent\textbf{Data availability.} Reduced data supporting the route-survival analyses are available from Zenodo \citep{kalathoor2026routeSurvivalDataV3}. The archive contains CSV source data for route metrics, contact and active-width histories, spectral partitions, mean-flow response, and selected three-dimensional support checks, together with a manifest and checksums.

\noindent\textbf{Use of generative AI and AI-assisted technologies.} During preparation of this work, AI-assisted tools were used for language editing, code review, and manuscript organization. The author reviewed and edited the resulting material and takes full responsibility for the content of this article.

\bibliography{references}

\end{document}

%% file: tables/case_nomenclature.tex
\begin{tabular}{llll}
\toprule
Symbol & Initial roughness & Mean shear & Role \\
\midrule
\(L\) & Low-mode & None & Broad connecting endpoint \\
\(M\) & Mixed & None & Delayed mixed-route reference \\
\(M'\) & Mixed & None & Independent mixed-route seed replicate \\
\(H\) & High-annulus & None & Localized route-memory endpoint \\
\(M+U_s\) & Mixed & \(u_s(z)\) & Mean-shear intervention on \(M\) \\
\bottomrule
\end{tabular}

%% file: tables/route_components.tex
\begin{tabular}{lrrrrrrr}
\toprule
Case & \(R_c\) & \(C_p\) & Contact & Vertical & Area & Spectral & Probe \\
\midrule
\(H\) & 0.2356 & 0.4752 & 0.0000 & 0.5713 & 0.3473 & 0.8456 & 0.2823 \\
\(M'\) & 0.5144 & 0.5253 & 0.3250 & 0.7417 & 0.4608 & 0.3856 & 0.4988 \\
\(M\) & 0.7185 & 0.5655 & 0.8250 & 0.7567 & 0.4737 & 0.3777 & 0.3874 \\
\(L\) & 0.9625 & 0.8446 & 1.0000 & 1.0000 & 1.0000 & 0.7497 & 0.3326 \\
\bottomrule
\end{tabular}

%% file: tables/route_transport.tex
\begin{tabular}{lrrr}
\toprule
Case & Route connection & Final salt flux & Cumulative exchange \\
\midrule
\(H\) & 0.2356 & 0.09812 & 1.797 \\
\(M'\) & 0.5144 & 0.1582 & 2.530 \\
\(M\) & 0.7185 & 0.1595 & 2.600 \\
\(L\) & 0.9625 & 0.3009 & 6.438 \\
\bottomrule
\end{tabular}

%% file: tables/final_shear_response.tex
\begin{tabular}{llrrrrl}
\toprule
Group & Quantity & \(M+U_s\) & \(M\) & \(M'\) & Ratio & Response \\
\midrule
Reach & \(w\) active width & 123.447 & 121.563 & 115.057 & 1.015 & inside tolerance \\
Reach & Salinity active width & 106.325 & 106.154 & 105.983 & 1.002 & inside tolerance \\
Spectral & Broad fraction & 0.6727 & 0.6028 & 0.6264 & 1.116 & modified branch \\
Spectral & Intermediate fraction & 0.0801 & 0.1510 & 0.1434 & 0.530 & modified branch \\
Spectral & Short-wave fraction & 0.0827 & 0.0647 & 0.0670 & 1.278 & elevated \\
Timing & First \(w\) contact & 57.75 & 57.75 & 60.00 & -- & same as mixed \\
Timing & First salinity contact & 59.50 & 59.50 & -- & -- & same as mixed \\
\bottomrule
\end{tabular}

%% file: tables/selected_3d_checks.tex
\begin{tabular}{rrrrrr}
\toprule
Time & Flux proxy & Mean \(|w|\) & \(w_\mathrm{rms}\) & \(w\) width & Salinity width \\
\midrule
45.0 & 0.05559 & 0.21486 & 0.41548 & 64.891 & 58.042 \\
58.0 & 0.10623 & 0.48185 & 0.70129 & 124.474 & 102.045 \\
60.0 & 0.11629 & 0.53976 & 0.76118 & 131.836 & 108.893 \\
\bottomrule
\end{tabular}

%% file: tables/resolution_support.tex
\begin{tabular}{lrrrr}
\toprule
Field & Production tail & Ultra tail & Production \(k_{95}/k_N\) & Ultra \(k_{95}/k_N\) \\
\midrule
\(S\) & \(4.81\times10^{-4}\) & \(3.89\times10^{-5}\) & 0.2329 & 0.1735 \\
\(T\) & \(3.13\times10^{-10}\) & \(6.92\times10^{-11}\) & 0.0644 & 0.0483 \\
\(w\) & \(5.99\times10^{-8}\) & \(1.62\times10^{-9}\) & 0.0890 & 0.0668 \\
\bottomrule
\end{tabular}